\documentclass[11pt,a4paper]{article}
\pdfoutput=1
\usepackage{jcappub}

\usepackage[T1]{fontenc}
\usepackage{mathtools}

\begin{document}

\title{Enhancement of gravitational waves at Q-ball decay including non-linear density perturbations}
\author[a,b]{Masahiro Kawasaki}
\author[c]{and Kai Murai}
\affiliation[a]{ICRR, University of Tokyo, Kashiwa, 277-8582, Japan}
\affiliation[b]{Kavli IPMU (WPI), UTIAS, University of Tokyo, Kashiwa, 277-8583, Japan}
\affiliation[c]{Department of Physics, Tohoku University, Sendai, 980-8578, Japan}

\abstract{%
    The existence of a stochastic gravitational wave background is indicated by the recent pulsar timing array (PTA) experiments.
    We study the enhanced production of second-order gravitational waves from the scalar perturbations when the universe experiences a transition from the early matter-dominated era to the radiation-dominated era due to Q-ball decay.
    We extend the analysis in previous work by including the frequency range where density perturbations go non-linear and find that the resultant gravitational wave spectrum can be consistent with that favored by the recent PTA experiment results.
}


\emailAdd{kawasaki@icrr.u-tokyo.ac.jp}
\emailAdd{kai.murai.e2@tohoku.ac.jp}

\begin{flushright}
    TU-1208
\end{flushright}

\maketitle

\section{Introduction}
\label{sec: intro}

Recently, the pulsar timing array (PTA) experiments, NANOGrav~\cite{NANOGrav:2023gor}, EPTA~\cite{Antoniadis:2023ott} with InPTA data, PPTA~\cite{Reardon:2023gzh}, and CPTA~\cite{Xu:2023wog}, have reported the signal of a stochastic gravitational wave background in the nHz range.
While it may indicate inspiraling supermassive black hole binaries~\cite{NANOGrav:2023hfp,Ellis:2023dgf,Broadhurst:2023tus,Depta:2023qst,Bi:2023tib,Zhang:2023lzt,Gouttenoire:2023nzr,Cannizzaro:2023mgc},%
\footnote{%
For earlier discussions on stochastic gravitational waves from inspiraling supermassive black hole binaries, see, e.g., Refs.~\cite{Rajagopal:1994zj,Jaffe:2002rt,Sesana:2008mz,Ravi:2012bz}.
}
various possibilities for its origin have been discussed such as cosmic strings~\cite{Ellis:2023tsl,Wang:2023len,Kitajima:2023vre,Eichhorn:2023gat,Lazarides:2023ksx,Servant:2023mwt,Antusch:2023zjk,Buchmuller:2023aus,Yamada:2023thl,Lazarides:2023rqf,Ahmed:2023rky,Maji:2023fhv},
domain walls~\cite{Lazarides:2023ksx,Guo:2023hyp,Kitajima:2023cek,Bai:2023cqj,Blasi:2023sej,Gouttenoire:2023ftk,Barman:2023fad,Lu:2023mcz,Du:2023qvj,Li:2023tdx,Babichev:2023pbf,Gelmini:2023kvo,Ge:2023rce,Zhang:2023nrs},
a first-order phase transition~\cite{Zu:2023olm,Han:2023olf,Megias:2023kiy,Fujikura:2023lkn,Addazi:2023jvg,Xiao:2023dbb,Li:2023bxy,Wang:2023bbc,Ghosh:2023aum,Cruz:2023lnq,DiBari:2023upq,Gouttenoire:2023bqy,Salvio:2023ynn,Ahmadvand:2023lpp,An:2023jxf},
second-order effects of scalar perturbations~\cite{Franciolini:2023pbf,Cai:2023dls,Inomata:2023zup,Wang:2023ost,Liu:2023ymk,Ebadi:2023xhq,Abe:2023yrw,Yi:2023mbm,Zhu:2023faa,Firouzjahi:2023lzg,You:2023rmn,HosseiniMansoori:2023mqh,Cheung:2023ihl,Balaji:2023ehk,Basilakos:2023xof,Jin:2023wri,Das:2023nmm,Zhao:2023joc,Liu:2023pau,Yi:2023tdk,Frosina:2023nxu,Bhaumik:2023wmw,Choudhury:2023wrm},
primordial tensor perturbations~\cite{Vagnozzi:2023lwo,Datta:2023vbs,Chowdhury:2023opo,Choudhury:2023kam,Gorji:2023sil,Ben-Dayan:2023lwd,Jiang:2023gfe,Zhu:2023lbf},
and other scenarios~\cite{Yang:2023aak,Li:2023yaj,Oikonomou:2023qfz,Borah:2023sbc,Murai:2023gkv,Niu:2023bsr,Unal:2023srk,Geller:2023shn,Bari:2023rcw}.
See also Refs.~\cite{Madge:2023cak,NANOGrav:2023hvm,EPTA:2023xxk,Lambiase:2023pxd,Franciolini:2023wjm,Bian:2023dnv,Figueroa:2023zhu,Wu:2023hsa,Ye:2023xyr,Cui:2023dlo,Ellis:2023oxs} for model comparison.

The existence of the scalar perturbations has been confirmed through the cosmic microwave background (CMB) and large-scale structure observations.
On the other hand, the primordial tensor perturbations have not been discovered on the CMB scales.
In addition to the quantum fluctuations during inflation, the tensor perturbations can be sourced by the second-order effects of the scalar perturbations.
In particular, if the universe experiences a matter-dominated era that suddenly transits to the standard radiation-dominated era, the scalar perturbations start to oscillate at the transition, and then gravitational waves can be abundantly generated~\cite{Inomata:2019ivs}.
This effect is called the ``poltergeist effect''.

The poltergeist effect has been studied in various contexts such as PBH evaporation~\cite{Inomata:2020lmk,Domenech:2020ssp,Domenech:2021wkk,Bhaumik:2022pil,Papanikolaou:2022chm,Bhaumik:2022zdd,Domenech:2023mqk,Basilakos:2023xof,Bhaumik:2023wmw}, Q-ball decay~\cite{White:2021hwi,Kasuya:2022cko}, oscillon decay~\cite{Lozanov:2022yoy}, and axion dynamics~\cite{Harigaya:2023ecg}.
Among them, we studied the poltergeist effect in the Q-ball decay linking to lepton asymmetry generation~\cite{Kawasaki:2022hvx} motivated by the recent determination of the primordial helium abundance~\cite{Matsumoto:2022tlr,Burns:2022hkq}.
In Ref.~\cite{Kasuya:2022cko}, we included the effect of the finite decay rate and pointed out that the generated gravitational waves can be probed by the future PTA experiment, SKA.

Previous works on the poltergeist effect have been limited in the frequency range due to non-linear density perturbations in a high-frequency region. 
However, we naturally expect that gravitational waves are significantly generated in such a region.
In this paper, we extend the analysis of the poltergeist effect using the fitting formula for non-linear density perturbations obtained from $N$-body simulations.
Even if the density perturbations exceed unity and become non-linear, the gravitational potential, which sources the gravitational waves, is related to the density perturbation through the Poisson equation and can remain linear.
By extending the frequency range, we find that the gravitational waves enhanced due to the Q-ball decay can be consistent with those favored by the PTA experiments and within reach of future experiments.

The rest of this paper is organized as follows.
In Sec.~\ref{sec: Q-ball scenario}, we briefly review the Q-ball scenario generating the lepton asymmetry mainly focusing on Q-ball decay.
The evolution of the scalar perturbations and production of gravitational waves in the Q-ball scenario including the non-linear regime are studied in Sec.~\ref{sec: GW}.
Our results are summarized and discussed in Sec.~\ref{sec: summary and discussion}.

\section{Q-ball scenario}
\label{sec: Q-ball scenario}

First, we briefly review the Q-ball scenario following Refs.~\cite{Kawasaki:2022hvx,Kasuya:2022cko}.
A Q-ball is a non-topological soliton often formed in the Affleck-Dine (AD) mechanism in the minimal supersymmetric standard model (MSSM)~\cite{Affleck:1984fy,Dine:1995kz}.
In the AD mechanism, a flat direction with a global U(1) charge ($=$ baryon or lepton number) starts to oscillate with a velocity in the phase direction and generates the charge in the early universe.
Depending on the shape of the scalar potential, the flat direction can fragment into spherical objects called Q-balls, which is classically stable due to the conservation of the U(1) charge ~\cite{Coleman:1985ki,Kusenko:1997zq,Kusenko:1997si,Enqvist:1997si,Kasuya:1999wu}.
Then, most of the charge generated in the AD mechanism is confined in the Q-balls.
In this paper, we focus on a delayed-type Q-ball~\cite{Kasuya:2001hg}.

\begin{figure}[t]
    \centering
    \includegraphics[width=.6\textwidth]{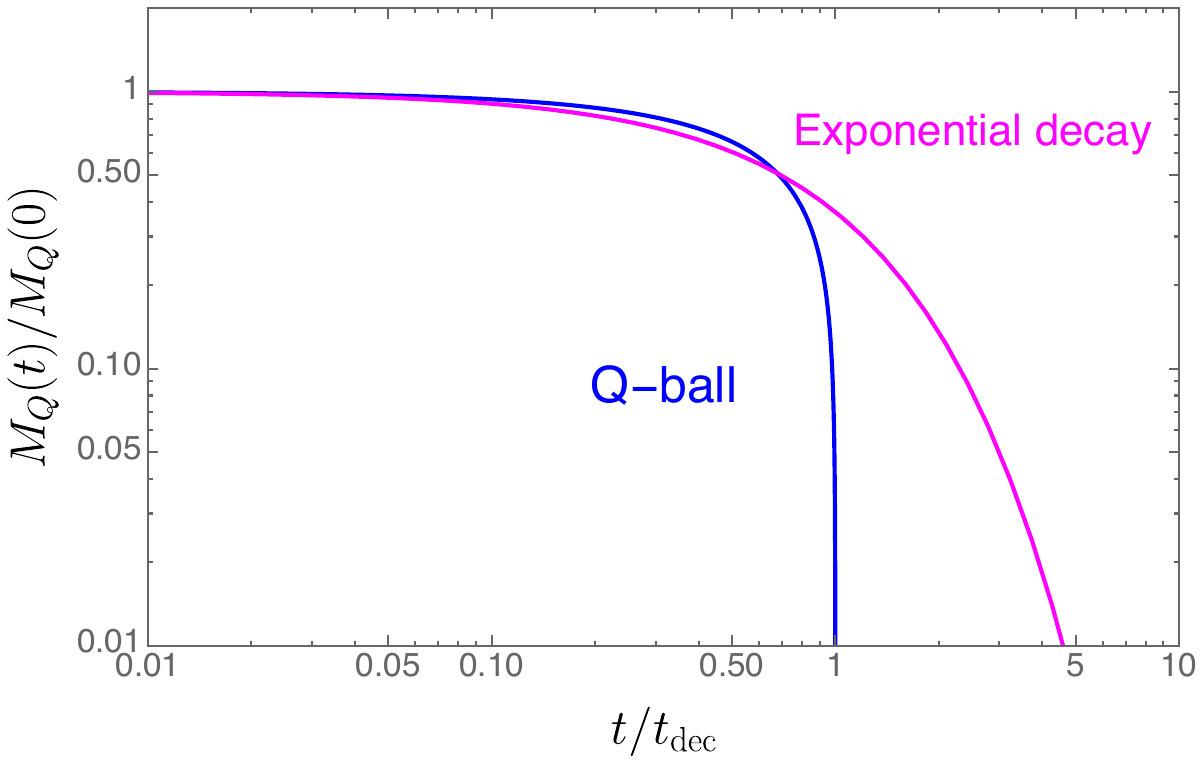}
    \caption{%
        Evolution of a Q-ball mass (blue) compared to the exponential decay (magenta) given by $M_Q(t) = M_Q(0)\exp(-t/t_\mathrm{dec})$.
    }
    \label{fig: Qball_decay}
    \end{figure}

Depending on the type of Q-ball and the U(1) charge that Q-balls carry, Q-balls can decay and release their charge.
For example, Q-balls with lepton charge decay emitting neutrinos.
In this case, the mass of each delayed-type Q-ball, $M_Q$, evolves as
\begin{align}
    M_Q(t) 
    =
    M_Q(0)
    \left( 1 - \frac{t}{ t_\mathrm{dec} } \right)^{3/5}
    \ ,
    \label{eq: Q-ball mass evolution}
\end{align}
where $t$ is the cosmic time, and $t = 0$ corresponds to the Q-ball formation.
$t_\mathrm{dec} (\propto M_Q(0)^{5/3})$ is the decay time, whose proportionality constant is determined by the model parameters.
In Fig.~\ref{fig: Qball_decay}, we show the evolution of the Q-ball mass following Eq.~\eqref{eq: Q-ball mass evolution} and that in the case of the exponential decay for comparison.
In the Q-ball decay, $M_Q$ decreases faster than the exponential decay, which is important for the poltergeist effect.

\begin{figure}[t]
    \centering
    \includegraphics[width=.75\textwidth]{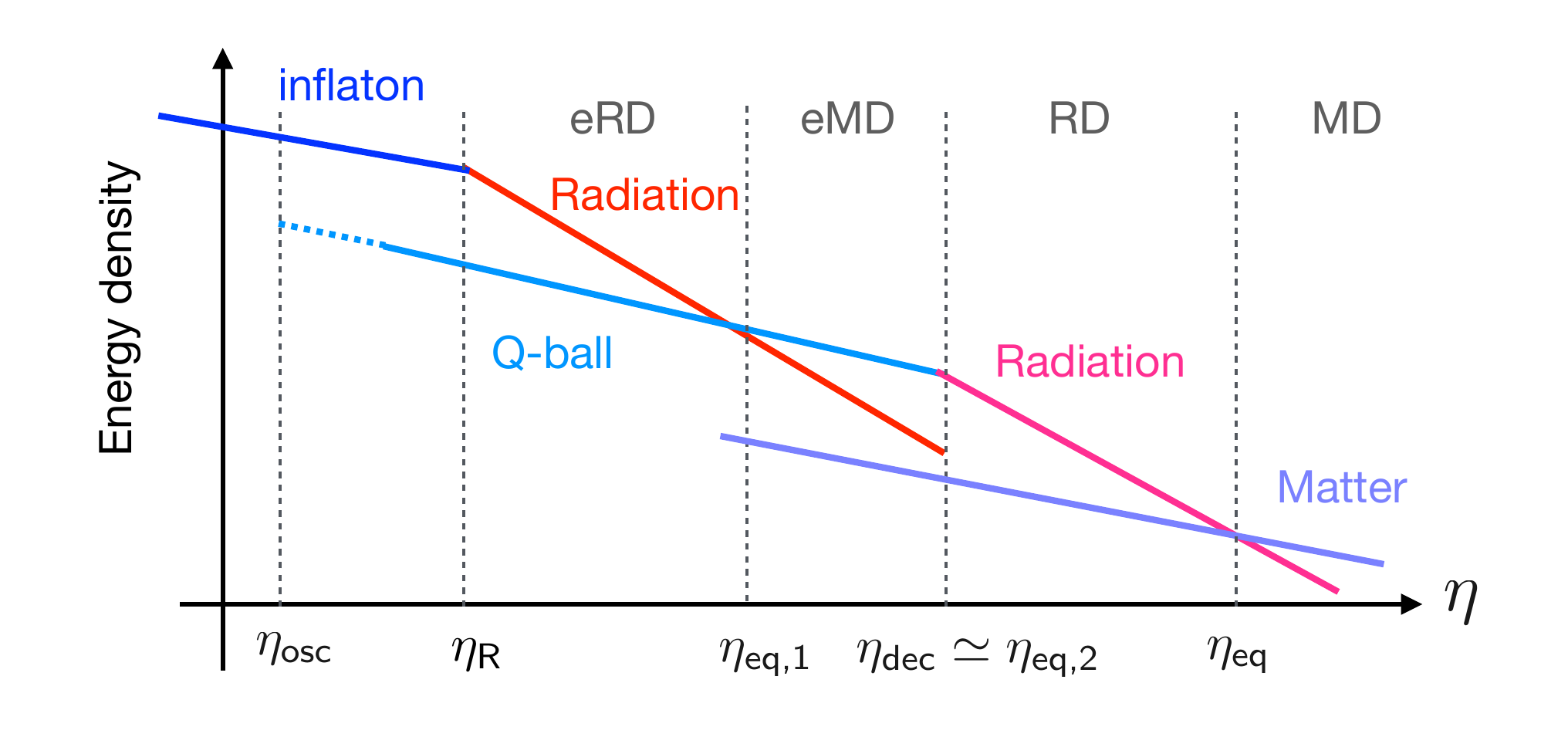}
    \caption{%
        Evolution of the energy densities of the inflaton, radiation from the inflaton decay, Q-balls, radiation from the Q-ball decay, and non-relativistic matter in our scenario.
        $\eta_\mathrm{osc}$ denotes the conformal time when the AD field starts oscillations.
        $\eta_\mathrm{R}$ represents the completion of the reheating.
        $\eta_\mathrm{eq,1}$ and $\eta_\mathrm{eq,2}$ are the conformal times at matter-radiation equality between radiation and Q-balls.
        $\eta_\mathrm{eq}$ is the standard matter-radiation equality time.
        }
    \label{fig: Thermal_history}
    \end{figure}

Since Q-balls behave as non-relativistic matter, they can dominate the universe until their decay.
In Fig.~\ref{fig: Thermal_history}, we show the time evolution of the energy densities of the inflaton, radiation from the inflaton decay, Q-balls, radiation from the Q-ball decay, and non-relativistic matter in our scenario.
The time coordinate is written in terms of the conformal time, $\eta$.
Just after the decay of the inflaton, the universe is dominated by radiation, which we call the early radiation-dominated (eRD) era.
Then, the Q-balls increase their energy fraction and come to have the energy density equal to that of radiation at $\eta = \eta_\mathrm{eq,1}$.
From then, the universe is in the early matter-dominated (eMD) era, which ends around the Q-ball decay at $\eta = \eta_\mathrm{dec}$.
We denote the equality time around the Q-ball decay by $\eta_\mathrm{eq,2}$, which is approximately equal to $\eta_\mathrm{dec}$.
Then, the standard radiation-dominated (RD) era succeeds in for $\eta > \eta_\mathrm{eq,2}$, which is followed by the standard matter-dominated (MD) era for $\eta > \eta_\mathrm{eq}$.

Now, we are interested in the enhancement of the second-order gravitational waves sourced by the scalar perturbations around the transition from the eMD era to the RD era.
To evaluate the second-order gravitational waves, we parameterize the relevant properties of the Q-balls by two quantities: the decay temperature $T_\mathrm{dec}$ corresponding to $t_\mathrm{dec}$ or $\eta_\mathrm{dec}$ and the energy density ratio $r_\mathrm{dec}$ of the Q-balls to radiation at $T = T_\mathrm{dec}$.
The decay temperature is related to the corresponding wavenumber that reenters the horizon at $T = T_\mathrm{dec}$ as
\begin{align}
    k_\mathrm{dec}
    \equiv 
    a_\mathrm{dec} H_\mathrm{dec}
    =
    1.16 \times 10^4 \, \mathrm{Mpc}^{-1}
    \frac{T_\mathrm{dec}}{1\,\mathrm{MeV}}
    \ ,
\end{align}
where we assumed $g_* (T_\mathrm{dec}) = g_{*s} (T_\mathrm{dec}) = 10.75$ and used $g_* (T_\mathrm{eq}) = 2 + 21/4 \times (4/11)^{4/3}$, $g_{*s} (T_\mathrm{eq}) = 43/11$, $a_\mathrm{eq} H_\mathrm{eq} = 0.0103~\mathrm{Mpc}^{-1}$~\cite{Planck:2018vyg}, and $T_\mathrm{eq} = 3400 \times 2.725~\mathrm{K}$. 
On the other hand, the energy density ratio, $r_\mathrm{dec}$, determines the duration of the eMD era through $\eta_\mathrm{dec}/\eta_\mathrm{eq,1} \sim \sqrt{r_\mathrm{dec}}$.
For example, $T_\mathrm{dec} = \mathcal{O}(1\,\text{--}\,10)$\,MeV and $r_\mathrm{dec} = \mathcal{O}(10^7)$ are obtained in Ref.~\cite{Kawasaki:2022hvx} while they are not the unique values to explain the lepton number generation.

In the eMD and RD eras ($\eta_\mathrm{eq, 1} \le \eta < \eta_\mathrm{eq}$), the scale factor, $a$, and conformal Hubble parameter, $\mathcal{H}$, are given respectively by
\begin{align}
    \frac{ a(\eta) }{ a(\eta_\mathrm{eq,1}) }
    &=
    \left\{
        \begin{array}{ll}
            \left( 
                \frac{\eta}{\eta_*}
            \right)^2
            + 
            \frac{2\eta}{\eta_*}
            \quad&
            (\eta \leq \eta_\mathrm{dec})
        \vspace{3mm} \\ 
            \frac{2\eta (\eta_\mathrm{dec} + \eta_*) - \eta_\mathrm{dec}^2}
            {\eta_*^2}
            \quad&
            (\eta > \eta_\mathrm{dec})
        \end{array}
    \right.
    \ ,
    \\[0.5em]
    \mathcal{H}(\eta)
    \equiv
    \frac{1}{ a(\eta) }
    \frac{ \mathrm{d} a(\eta) }{ \mathrm{d} \eta }
    &=
    \left\{
        \begin{array}{ll}
            \frac{2(\eta + \eta_*)}{\eta^2 + 2\eta\eta_*}
            \hspace{16.2mm}&
            (\eta \leq \eta_\mathrm{dec})
        \vspace{3mm} \\ 
            \frac{1}{\eta - \frac{\eta_\mathrm{dec}^2}{2(\eta_\mathrm{dec} + \eta_*)}}
            &
            (\eta > \eta_\mathrm{dec})
        \end{array}
    \right.
    \ ,
\end{align}
where $\eta_* \equiv \eta_\mathrm{eq,1}/(\sqrt{2} - 1)$.
Here, we assumed sudden transitions at $\eta_\mathrm{eq,1}$ and $\eta_\mathrm{dec} \simeq \eta_\mathrm{eq,2}$.

\section{Gravitational wave generation in the Q-ball scenario}
\label{sec: GW}

In this section, we discuss the enhancement of the second-order gravitational waves around the transition from the eMD era to the RD era.
To this end, we first consider the evolution of the gravitational potential and then evaluate the sourced gravitational waves.

\subsection{Scalar perturbations}
\label{subsec: scalar}

Here, we discuss the evolution of the gravitational potential around the eMD era.
We adopt the conformal Newtonian gauge and consider the metric perturbations given by
\begin{align}
    \mathrm{d} s^2
    =
    a^2 
    \left[ 
        - ( 1 + 2\Phi ) \mathrm{d} \eta^2
        + 
        \left(
            ( 1 - 2\Psi ) \delta_{i j} + \frac{1}{2} h_{i j}
        \right)
        \mathrm{d} x^i \mathrm{d} x^j
    \right]
    \ ,
\end{align}
where $\Phi$ and $\Psi$ are scalar perturbations, and $h_{i j}$ is tensor perturbations.
In the early universe, we can neglect the anisotropic stress and obtain $\Phi = \Psi$.\footnote{
The anisotropic stress may be produced when the density perturbations go non-linear. Here we assume that the induced anisotropic stress is negligible.}
To evaluate the production of gravitational waves, the evolution of the gravitational potential, $\Phi$, is a key ingredient.
Thus, we focus on the behavior of the transfer function, $\mathcal{T}$, of $\Phi$ around the eMD era.
In particular, we discuss three effects: the transition from the eRD era to the eMD era, that from the eMD era to the RD era, and the non-linear evolution of the density perturbations during the eMD era.
While the first two effects have been discussed in previous work~\cite{Inomata:2020lmk,Kasuya:2022cko}, the last one is the new effect that we first include in this paper.

First, we consider the transition from the eRD era to the eMD era.
Around this transition, it is sufficient to discuss the scalar perturbations in the linear regime.
In the eMD era, the gravitational potential becomes constant in the linear regime as in the standard MD era.
We express this constant value of the transfer function by $\mathcal{T}_\mathrm{plateau}$.
While $\Phi$ on superhorizon scales at the transition is not suppressed at the transition, $\Phi$ on subhorizon scales at the transition is suppressed during the eRD era.
The fitting formula of $\mathcal{T}_\mathrm{plateau}$ is given by~\cite{Bardeen:1985tr,Inomata:2020lmk}
\begin{align}
    \mathcal{T}_\mathrm{plateau}(x_\mathrm{eq,1})
    \equiv &
    \mathcal{T}(x)|_{ \eta_\mathrm{eq,1} \ll \eta \lesssim \eta_\mathrm{dec} }
    \nonumber
    \\[0.5em]
    \simeq &
    \frac{ \ln[1 + 0.146 x_\mathrm{eq,1}] }{ 0.146 x_\mathrm{eq,1} }
    \nonumber\\
    & \times 
    \left[ 
        1 + 0.242 x_\mathrm{eq,1}
        + \left( 1.01  x_\mathrm{eq,1}\right)^2
        + \left( 0.341 x_\mathrm{eq,1}\right)^3
        + \left( 0.418 x_\mathrm{eq,1}\right)^4
    \right]^{-1/4}
    \ ,
\end{align}
where $x_\mathrm{eq,1} \equiv k \eta_\mathrm{eq,1}$ and $x \equiv k \eta$ with $k$ being the wavenumber of $\Phi$.
Note that, here, we normalize $\mathcal{T}_\mathrm{plateau}$ so that $\mathcal{T}_\mathrm{plateau}(x_\mathrm{eq,1} \to 0) \to 1$.
In the limit of $x_\mathrm{eq,1} \gg 1$, the constant value is suppressed as $\mathcal{T}_\mathrm{plateau}(x_\mathrm{eq,1} \gg 1) \propto x_{\mathrm{eq},1}^{-2}\ln x_{\mathrm{eq},1}$.

Next, we consider the transition from the eMD era to the RD era.
Around this transition, $\mathcal{T}$ decays proportionally to the matter density perturbations at first.
In this phase, $\mathcal{T}$ approximately decays as~\cite{Inomata:2019zqy}
\begin{align}
    \frac{\mathcal{T}(t)}{\mathcal{T}_\mathrm{eMD}}
    \simeq
    \frac{M_Q(t)}{M_{Q,\mathrm{init}}}
    \simeq
    \left( 1 - \frac{t}{t_\mathrm{dec}} \right)^{3/5}
    \ ,
    \label{eq: transfer in Q-ball decay}
\end{align}
where $\mathcal{T}_\mathrm{eMD}$ is the transfer function just before the Q-ball decay.
While $\mathcal{T}_\mathrm{eMD} = \mathcal{T}_\mathrm{plateau}$ in the linear regime, it can be different from $\mathcal{T}_\mathrm{plateau}$ if the density perturbations become non-linear as discussed below.
The time evolution of Eq.~\eqref{eq: transfer in Q-ball decay} assumes that $\Phi (\propto \mathcal{T})$ is determined only by the matter density fluctuation, which requires
\begin{align}
    &3 a^2 |\ddot{\mathcal{T}}|
    \ll
    k^2 \mathcal{T}
    \label{eq: decouple condition}
    \\
    \Leftrightarrow \quad &
    \frac{18}{25 (t_\mathrm{dec} - t)^2 } \ll \frac{k^2}{a^2}
    \ ,
    \label{eq: T-delta proportionality condition}
\end{align}
as a necessary condition.
Once this condition is violated, $\Phi$ decouples from the matter density perturbations and starts to independently oscillate.%
\footnote{%
    Since Eq.~\eqref{eq: T-delta proportionality condition} is a necessary condition, $\Phi$ can decouple from the matter density perturbations even before this condition is violated.
    Thus, the violation of Eq.~\eqref{eq: T-delta proportionality condition} gives a lower bound for $\mathcal{T}$.
}
To quantify this effect, we define the suppression factor of $\mathcal{T}$ at the onset of oscillations by $S$ satisfying $S(k) = \mathcal{T}_\mathrm{eMD}$ in the sudden decay limit.
From $\eta a = 3t$ during the eMD era, we obtain the decoupling time $\eta_\mathrm{dcpl}$ for a given $k$ as
\begin{align}
    k \eta_\mathrm{dec} - k \eta_\mathrm{dcpl}
    \simeq
    \frac{9\sqrt{2}}{5}
    \ ,
\end{align}
which leads to the lower bound of $S$, $S_\mathrm{low}$, given by
\begin{align}
    S_\mathrm{low}
    \simeq 
    \left( \frac{9\sqrt{2}}{5 k \eta_\mathrm{dec}}\right)^{3/5}
    \mathcal{T}_\mathrm{eMD}
    \equiv 
    S_\mathrm{dec} \mathcal{T}_\mathrm{eMD}
    \ .
    \label{eq: S lower bound}
\end{align}
Here, $S_\mathrm{dec}$ denotes the suppression around the transition from the eMD era to the RD era.
Note that this estimate of $S_\mathrm{dec}$ uses the equation of motion for $\Phi$ in the linear regime, and how the gravitational potential in the non-linear regime decays is non-trivial.
However, Eq.~\eqref{eq: transfer in Q-ball decay} is derived from the Poisson equation, which is expected to be valid even in the non-linear regime.
Furthermore, as the Q-ball decay proceeds, the non-linearity decreases, which may partly justify the use of Eqs.~\eqref{eq: decouple condition} and \eqref{eq: T-delta proportionality condition}.
Therefore, in the following, we assume that the estimate above can also be applied to the non-linear regime.

After the gravitational potential decouples from the matter density fluctuation, the transfer function in the RD era follows the equation of 
\begin{align}
    \mathcal{T}'' + 4 \mathcal{H} \mathcal{T}' + \frac{k^2}{3} \mathcal{T}
    =
    0
    \ ,
\end{align}
where the primes denote the derivatives with respect to $\eta$.
We can obtain $\mathcal{T}$ after the decoupling by solving this equation with the approximated initial conditions, $\mathcal{T}(x_\mathrm{dec} \equiv k\eta_\mathrm{dec}) = S_\mathrm{low}(k)$ and $\mathcal{T}'(x_\mathrm{dec}) = 0$.

Finally, we discuss the non-linear evolution of the scalar perturbations.
For simplicity, in this paper, we adopt the scale-invariant spectrum for the primordial curvature perturbations:
\begin{align}
    \mathcal{P}_\zeta (k)
    =
    C^2 A_\mathrm{s}
    \ ,
    \label{eq: Pzeta}
\end{align}
where $A_\mathrm{s} = 2.1 \times 10^{-9}$ is the amplitude of the power spectrum on the CMB pivot scale $k_* = 0.05~\mathrm{Mpc}^{-1}$~\cite{Planck:2018vyg}.
Here, we introduced a constant $C$ because we are focusing on scales much smaller than the CMB scale, and $\mathcal{P}_\zeta$ can be larger than $A_s$ on such scales.
Such an enhancement of the curvature perturbations can be achieved, for instance, by modifying the inflaton potential in the single-field slow-roll inflation~\cite{Starobinsky:1992ts,Ivanov:1994pa} or by considering multi-stage inflation~\cite{Silk:1986vc}.
For $k \gtrsim 10^5\,\mathrm{Mpc}^{-1}$, the curvature perturbation is bounded as $\mathcal{P}_\zeta < \mathcal{O}(10^{-2})$ ($C < \mathcal{O}(10^3)$) from the PBH formation~\cite{Gow:2020bzo} (see also Ref.~\cite{FrancoAbellan:2023sby} for other constraints on the curvature perturbations).
In this paper, we assume that Eq.~\eqref{eq: Pzeta} is applicable at least for $k \gtrsim 10^5\,\mathrm{Mpc}^{-1}$ and evaluate the gravitational wave using it even for all $k$.
Then, this approximation is valid for the gravitational wave power spectrum on $k \gtrsim 10^6\,\mathrm{Mpc}^{-1}$, which is of main interest in the context of the PTA signals, while that on larger scales may be suppressed compared to our result.

During the eMD era, the gravitational potential is almost constant in the linear regime.
The linear density perturbation, $\delta_\mathrm{m,L}$, can be evaluated from the Poisson equation as
\begin{align}
    \frac{3}{5} k^2 \mathcal{T}_\mathrm{plateau} (x_\mathrm{eq,1}) 
    \mathcal{P}_\zeta^{1/2}
    = 
    \frac{3}{2} \mathcal{H}^2 \delta_\mathrm{m,L}(k)
    \ ,
    \label{eq: Poisson in linear}
\end{align}
where the factor of $3/5$ comes from the relation between the gravitational potential and the curvature perturbation during the eMD era, $\Phi = (3/5) \mathcal{T} \zeta$.
Since $\mathcal{H} \propto a^{-1/2}$ in the eMD era, the density perturbation grows as $\delta_\mathrm{m,L} \propto a$.
As a result, the density perturbation becomes non-linear during the eMD era for large $k$.

\begin{figure}[t]
    \centering
    \includegraphics[width=.8\textwidth]{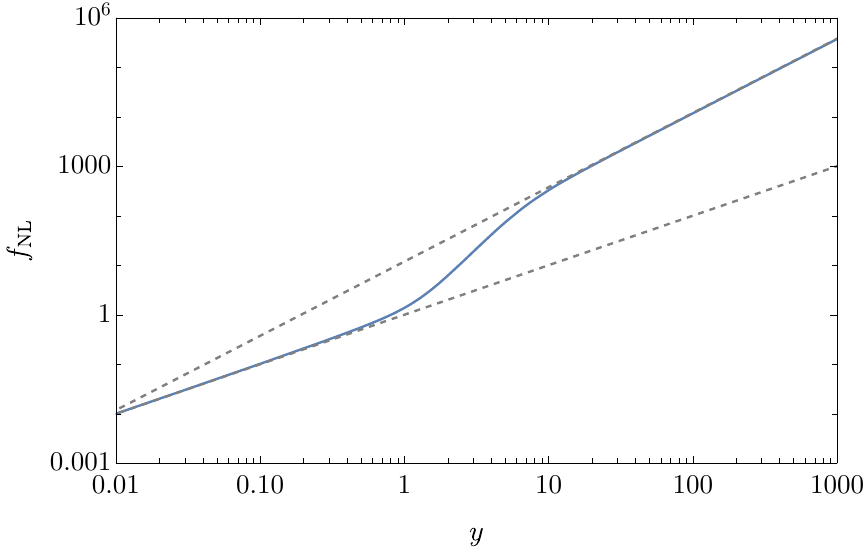}
    \caption{%
        Fitting function of non-linear density perturbations, $f_\mathrm{NL}(y)$.
        The upper and lower gray dashed lines denote $11.86 y^{3/2}$ and $y$, respectively.
    }
    \label{fig: fNL}
    \end{figure}

\begin{figure}[t]
    \centering
    \includegraphics[width=.8\textwidth]{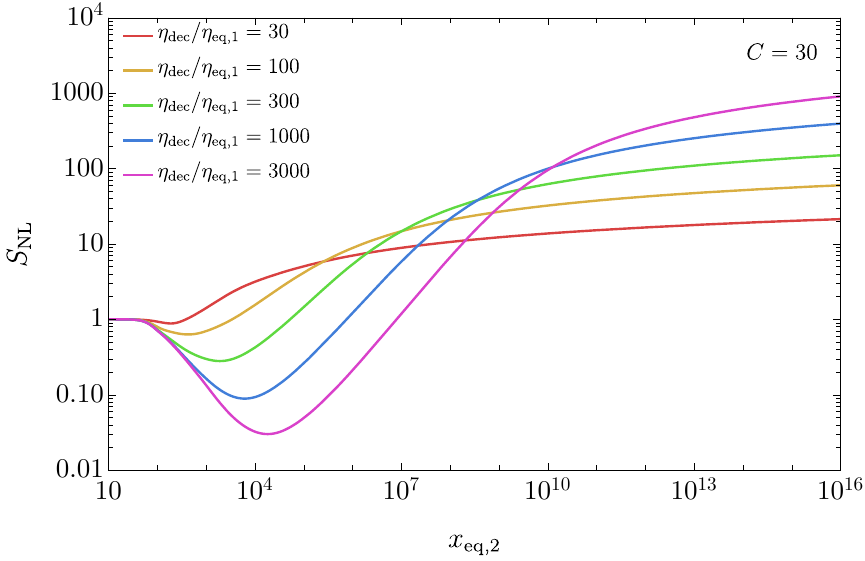}
    \caption{%
        $S_\mathrm{NL}$ for $C = 30$ and $\eta_\mathrm{dec}/\eta_\mathrm{eq,1} = 30, 100, 300, 1000$, and $3000$.
    }
    \label{fig: SNL}
    \end{figure}

Once the density perturbation becomes non-linear, the evolution in the linear regime is no longer applicable, and the gravitational potential evolves in time.
From the $N$-body simulations, the non-linear density fluctuation, $\delta_\mathrm{m,NL}$ can be related to the linearly extrapolated density fluctuations $\delta_\mathrm{m,L}$ using a function $f_\mathrm{NL}$ as~\cite{Hamilton:1991es,Peacock:1993xg}
\begin{align}
    \delta_\mathrm{m,NL}^2 (k_\mathrm{NL})
    =
    f_\mathrm{NL}[ \delta_\mathrm{m,L}^2 (k_\mathrm{L}) ]
    \ ,
    \label{eq: fNL definition}
\end{align}
where $k_\mathrm{L}$ is related to $k_\mathrm{NL}$ as
\begin{align}
    k_\mathrm{L} 
    =
    \left[ 1 + \delta_\mathrm{m,NL}^2 (k_\mathrm{NL}) \right]^{-1/3} 
    k_\mathrm{NL}
    \ .
    \label{eq: kL vs kNL}
\end{align}
In this paper, we adopt a fitting formula of $f_\mathrm{NL}$ given in Ref.~\cite{Peacock:1993xg}:%
\footnote{%
    There are more recent results on the non-linear evolution of the density perturbations such as Refs.~\cite{Mo:1995db,Peacock:1996ci,Smith:2002dz,Takahashi:2012em,Mead:2015yca,Mead:2020vgs}.
    However, they mainly focus on the power spectrum for the red-tilted curvature perturbations or that in the $\Lambda$CDM model.
    Now, we are interested in the non-linear evolution of the density perturbations for a scale-invariant $\mathcal{P}_\zeta$ in the matter-dominated universe.
    Thus, we adopted the fitting formula in Ref.~\cite{Peacock:1993xg}, which agrees with the result for the correlation function including the case with a scale-invariant curvature perturbations~\cite{Hamilton:1991es}.
}
\begin{align}
    f_\mathrm{NL}(y)
    =
    y \left[ 
        \frac{1 + 0.4 y + 0.498 y^4}
        {1 + 0.00365 y^3}
    \right]^{1/2}
    \ ,
\end{align}
where $\delta_\mathrm{m,L}^2 (k_\mathrm{L})$ is substituted to $y$ in Eq.~\eqref{eq: fNL definition}.
Since this formula is obtained from a fitting to $N$-body data, it can be inaccurate for the region where the fitted data is absent.
In this paper, we consider that $f_\mathrm{NL}$ is reliable at least $f_\mathrm{NL} < 10^3$ from the result in Ref.~\cite{Hamilton:1991es}.
In the limit of $y \gg 1$, we obtain $f_\mathrm{NL} \propto y^{3/2}$ while we obtain $f_\mathrm{NL} \simeq y$ for $y \ll 1$.
We can understand this dependency for $y \gg 1$ from $\delta_\mathrm{m,NL}^2 \propto a^{3}$ in the stable clustering and $\delta_\mathrm{m,L}^2 \propto a^{2}$ in the linear growth~\cite{Hamilton:1991es}.
We show $f_\mathrm{NL}(y)$ in Fig.~\ref{fig: fNL}.
Even in the non-linear regime, the gravitational potential is related to the density fluctuation through the Poisson equation~\cite{Bagla:1995ya}.
We define the correction factor of the gravitational potential in the non-linear regime compared to the linear solution, $S_\mathrm{NL}$, by
\begin{align}
    \frac{3}{5} k^2 
    S_\mathrm{NL}(k)
    \mathcal{T}_\mathrm{plateau} (x_\mathrm{eq,1}) 
    \mathcal{P}_\zeta^{1/2}
    = 
    \frac{3}{2} \mathcal{H}^2 \delta_\mathrm{m,NL}(k)
    \ .
\end{align}
Then, $\mathcal{T}_\mathrm{eMD}$ is represented as $\mathcal{T}_\mathrm{eMD} = S_\mathrm{NL} \mathcal{T}_\mathrm{plateau}$.
We show $S_\mathrm{NL}(k)$ for $C = 30$ and several values of $\eta_\mathrm{dec}/\eta_\mathrm{eq,1}$ in Fig.~\ref{fig: SNL}.
While $f_\mathrm{NL}(x) > x$ for all $x$ as shown in Fig.~\ref{fig: fNL}, $S_\mathrm{NL}$ can be less than unity because $f_\mathrm{NL}$ relates the linear and non-linear density perturbations in different $k$ (see Eqs.~\eqref{eq: fNL definition} and \eqref{eq: kL vs kNL}).
Incorporating $S_\mathrm{NL}$, we define the correction factor for the gravitational potential due to the two transitions and non-linear effect, $S_\mathrm{low,NL}$, by
\begin{align}
    S_\mathrm{low,NL}
    \equiv 
    S_\mathrm{dec} S_\mathrm{NL}
    \mathcal{T}_\mathrm{plateau}
    \ .
    \label{eq: S lower bound NL}
\end{align}
In Appendix~\ref{Appendix}, we show the frequency dependence of $S_\mathrm{low,NL}$ and its ingredients for the model parameters adopted below.

\subsection{Second-order gravitational waves}
\label{subsec: 2nd GW}

Next, we consider the generation of the second-order gravitational waves from the gravitational potential.
We parameterize the gravitational wave energy density by the density parameter per $\ln k$ given by
\begin{align}
    \Omega_\mathrm{GW} (\eta, k)
    &\equiv
    \frac{ \rho_\mathrm{GW}(\eta, k) }{ \rho_\mathrm{tot}(\eta) }
    \nonumber\\
    &=
    \frac{1}{24} \left( \frac{ k }{ a(\eta) H(\eta) } \right)^2
    \overline{ \mathcal{P}_h (\eta, k) }
    \ ,
\end{align}
where $\rho_\mathrm{tot}$ is the total energy density at $\eta$, and $\overline{ \mathcal{P}_h (\eta, k) }$ is the time average of the gravitational wave power spectrum.
The power spectrum of the second-order gravitational waves is given by~\cite{Kohri:2018awv} (see also Refs.~\cite{Ananda:2006af,Baumann:2007zm,Saito:2008jc,Saito:2009jt,Bugaev:2009zh,Inomata:2016rbd})
\begin{align}
    \overline{ \mathcal{P}_h (\eta, k) }
    =
    4 \int_0^\infty \mathrm{d}v \int_{|1-v|}^{1+v} \mathrm{d}u \,
    \left[
        \frac{ 4v^2 - (1 + v^2 - u^2)^2 }{ 4 v u } 
    \right]^2
    \overline{ I^2(u, v, k, \eta, \eta_\mathrm{dec}) }
    \mathcal{P}_\zeta(u k) \mathcal{P}_\zeta(v k)
    \ ,
\end{align}
where $I(u, v, k, \eta, \eta_\mathrm{dec})$ is expressed as
\begin{align}
    I(u, v, k, \eta, \eta_\mathrm{dec})
    =
    \int_0^x \mathrm{d} \tilde{x} \,
    \frac{ a(\tilde{\eta})}{ a(\eta) } k G_k(\eta, \tilde{\eta})
    f(u, v, \tilde{x}, x_\mathrm{dec})
    \ ,
\end{align}
with $\tilde{x} \equiv k\tilde{\eta}$.
Here, $G_k(\eta, \tilde{\eta})$ is the Green function representing the propagation of the gravitational waves sourced at $\tilde{\eta}$ and satisfies
\begin{align}
    G_k''(\eta, \tilde{\eta})
    + \left( k^2 - \frac{a''(\eta)}{a(\eta)} \right) G_k(\eta, \tilde{\eta})
    =
    \delta(\eta - \tilde{\eta})
    \ ,
\end{align}
where the primes denote the derivatives with respect to not $\tilde{\eta}$ but $\eta$.
$f(u, v, \tilde{x}, x_\mathrm{dec})$ represents the second-order source term of the gravitational waves and is given by
\begin{align}
    & f(u, v, \tilde{x}, x_\mathrm{dec})
    \nonumber\\
    & \equiv
    \frac{
        3
        \left[
            2 ( 5 + 3w(\tilde{\eta}) ) \mathcal{T}(u \tilde{x}) \mathcal{T}(v \tilde{x}) 
            + 4 \mathcal{H}^{-1} 
            ( \mathcal{T}'(u \tilde{x}) \mathcal{T}(v \tilde{x}) 
            + \mathcal{T}(u \tilde{x}) \mathcal{T}'(v \tilde{x}) )
            + 4 \mathcal{H}^{-2} \mathcal{T}'(u \tilde{x}) \mathcal{T}'(v \tilde{x}) 
        \right]
    }
    {25 ( 1 + w(\tilde{\eta}) )}
    \ ,
\end{align}
where $w(\tilde{\eta}) \equiv p/\rho$ is the equation-of-state parameter at $\tilde{\eta}$, and $\mathcal{T}(x)$ is an abbreviated notation of $\mathcal{T}(x, x_\mathrm{eq,1}, x_\mathrm{dec})$.
Note that the primes denote derivatives with respect to the conformal time as mentioned above, and thus $\mathcal{T}'(u \tilde{x}) = \partial \mathcal{T}(u \tilde{x})/\partial \tilde{\eta} = u k \partial \mathcal{T}(u \tilde{x})/\partial (u \tilde{x})$.

We evaluate the gravitational wave spectrum at a certain time, $\eta = \eta_c$, after the energy density of the gravitational waves becomes constant and before the late-time equality.
Then, we translate it into the current energy density spectrum.
Due to the resonance among the oscillations of the gravitational potentials in $f$ and the Green function $G_k$, $\overline{\mathcal{P}_h}$ has a resonant contribution.
In this paper, we focus on the dominant resonant contribution.

First, we review the evaluation with a cutoff of the scalar perturbations at $k = k_\mathrm{cut}$, which is defined by $\delta_\mathrm{m,L}(k_\mathrm{cut}) = 1$ in Eq.~\eqref{eq: Poisson in linear}.
With this cutoff, the gravitational wave spectrum is estimated as~\cite{Inomata:2019ivs,Inomata:2020lmk}
\begin{align}
    \Omega_{\mathrm{GW,RD}}^{(\mathrm{res})} (\eta_c, k)
    \simeq 
    2.9 \times 10^{-7} Y 
    \mathcal{P}_\zeta^2 S_\mathrm{low}^4(k)
    (k \eta_\mathrm{dec})^7 F(s_0(k), n_\mathrm{s,eff})
    \ ,
    \label{eq: 2nd GW formula}
\end{align}
where $Y \simeq 2.3$ is the numerical fudge factor, and $F(s_0(k), n_\mathrm{s,eff})$ reflects the effect of the cutoff at $k = k_\mathrm{cut}$ in the wavenumber integration.
$F(s_0(k), n_\mathrm{s,eff})$ is defined by
\begin{align}
    &F(s, n_\mathrm{s,eff})
    \nonumber \\
    &\equiv
    2 s \left( \frac{3}{4} \right)^{n_\mathrm{s,eff}-1}
    \nonumber \\
    & \phantom{\equiv} \times 
    \left[ 
        4 \,_2F_1 \left( \frac{1}{2}, 1-n_\mathrm{s,eff}; \frac{3}{2};\frac{s^2}{3}\right)
        -
        3 \,_2F_1 \left( \frac{1}{2}, -n_\mathrm{s,eff}; \frac{3}{2};\frac{s^2}{3}\right)
        -s^2 \,_2F_1 \left( \frac{3}{2}, -n_\mathrm{s,eff}; \frac{5}{2};\frac{s^2}{3}\right)
    \right]
    \ ,
\end{align}
and
\begin{align}
    s_0(k)
    =
    \left\{
        \begin{array}{ll}
            1
            \quad&
            (k \leq \frac{2}{1 + \sqrt{3}} k_\mathrm{cut})
        \vspace{3mm} \\ 
            2 \frac{k_\mathrm{cut}}{k} - \sqrt{3}
            \quad&
            (\frac{2}{1 + \sqrt{3}} k_\mathrm{cut} \leq k \leq \frac{2}{\sqrt{3}} k_\mathrm{cut})
        \vspace{3mm} \\ 
            0
            \quad&
            (k \geq \frac{2}{\sqrt{3}} k_\mathrm{cut})
        \end{array}
    \right.
    \ ,
\end{align}
with $_2F_1$ being a hypergeometric function.
Here, $n_\mathrm{s,eff}$ represents the effective spectral index of the scalar perturbations including the suppression effects discussed above, $\mathcal{T}_\mathrm{plateau}$ and $S_\mathrm{dec}$.
Note that this formula is obtained for a constant value of $n_\mathrm{s,eff}$.

Next, we extend this formula to include the nonlinear regime, $k > k_\mathrm{cut}$.
Specifically, we eliminate the cutoff at $k_\mathrm{cut}$ and introduce $S_\mathrm{NL}$. 
Since we do not set a cutoff of the scalar perturbations, we replace $s_0(k)$ by unity in Eq.~\eqref{eq: 2nd GW formula}.
Considering the discussion in Sec.~\ref{subsec: scalar}, we can decompose $n_\mathrm{s,eff}$ into four contributions as
\begin{align}
    n_\mathrm{s, eff}
    =
    n_s + \Delta n_\mathrm{s,plateau} + \Delta n_\mathrm{s,dec} 
    + \Delta n_\mathrm{s,NL}
    \ .
\end{align}
Here, $n_s = 1$ is the index of $\mathcal{P}_\zeta$ determined in Eq.~\eqref{eq: Pzeta},
$\Delta n_\mathrm{s,plateau}$ represents the momentum dependence of $\mathcal{T}_\mathrm{plateau}$ and is given by
\begin{align}
    \Delta n_\mathrm{s,plateau} (k)
    =
    2 \frac{\mathrm{d} \ln \mathcal{T}_\mathrm{plateau}}{\mathrm{d} \ln k}
    \simeq 
    \left\{
    \begin{array}{cc}
        0 \quad & (k \eta_\mathrm{eq, 1} \ll 1) \\
        -4 \quad & (k \eta_\mathrm{eq, 1} \gg 1)
    \end{array}
    \right.
    \ ,
\end{align}
$\Delta n_\mathrm{s,dec} = -6/5$ reflects the $k$-dependence of $S_\mathrm{dec}$ in Eq.~\eqref{eq: S lower bound},
and $\Delta n_\mathrm{s,NL}$ accounts for the non-linear evolution of $\Phi$ by
\begin{align}
    \Delta n_\mathrm{s,NL}(k)
    =
    2 \frac{\mathrm{d} \ln S_\mathrm{NL}}{\mathrm{d} \ln k}
    \ .    
\end{align}

\begin{figure}[t]
    \centering
    \includegraphics[width=.8\textwidth]{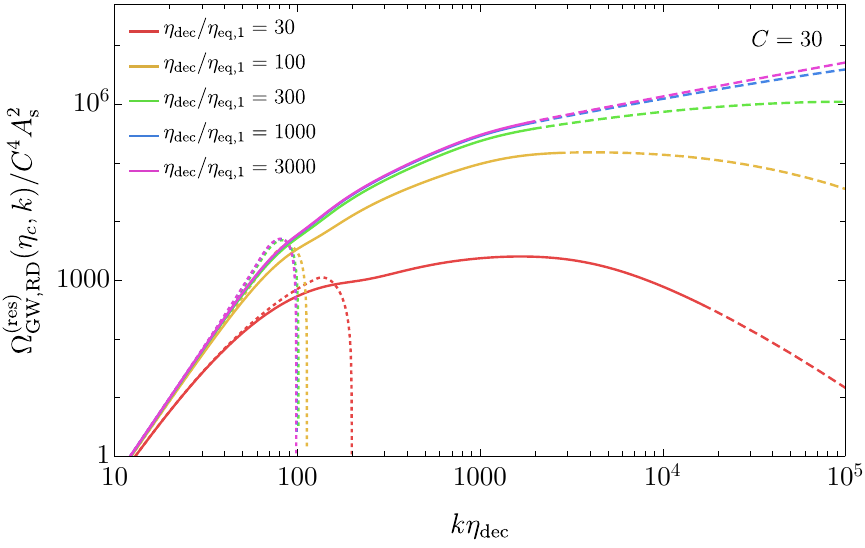}
    \caption{%
        Shapes of the gravitational wave spectrum for different values of $\eta_\mathrm{dec}/\eta_\mathrm{eq,1}$.
        Here, we fix $C = 30$.
        In the regions with dashed lines, the estimate of $S_\mathrm{NL}$ can be inaccurate.
        The dotted lines denote the gravitational wave spectrum from the linear scalar perturbations with the cutoff at $k = k_\mathrm{cut}$.
    }
    \label{fig: GW shape}
    \end{figure}

Although the formula of Eq.~\eqref{eq: 2nd GW formula} is obtained for a constant $n_\mathrm{s,eff}$, we substitute the $k$-dependent $n_\mathrm{s,eff}$ as a rough estimate.
Then, we obtain the gravitational wave spectrum as
\begin{align}
    \Omega_{\mathrm{GW,RD}}^{(\mathrm{res})} (\eta_c, k)
    \simeq 
    2.9 \times 10^{-7} Y 
    \mathcal{P}_\zeta^2 S_\mathrm{low,NL}^4(k)
    (k \eta_\mathrm{dec})^7
    F(1, n_\mathrm{s,eff}(k))
    \ .
    \label{eq: 2nd GW formula Mod}
\end{align}
This formula still depends on $\eta_\mathrm{dec}$ and $\eta_\mathrm{eq,1}$, which are determined by $T_\mathrm{dec}$ and $\eta_\mathrm{dec}/\eta_\mathrm{eq,1}$.
Thus, we can evaluate $\Omega_{\mathrm{GW,RD}}^{(\mathrm{res})}$ for fixed values of $T_\mathrm{dec}$, $\eta_\mathrm{dec}/\eta_\mathrm{eq,1}$, and $C$.
We show the shape of the gravitational wave abundance for $C = 30$ and different values of $\eta_\mathrm{dec}/\eta_\mathrm{eq,1}$ in Fig.~\ref{fig: GW shape}.
The smaller $\eta_\mathrm{dec}/\eta_\mathrm{eq,1}$ is, the more the gravitational waves are suppressed on large $k$ due to $\mathcal{T}_\mathrm{plateau}$ included in $S_\mathrm{low,NL}$.
The region where $\delta_\mathrm{m,NL}^2(k) > 10^3$ is shown by dashed lines.
For comparison, we also show the gravitational wave spectrum from the linear scalar perturbations with the cutoff at $k = k_\mathrm{cut}$ by dotted lines.

\begin{figure}[t]
    \centering
    \includegraphics[width=.8\textwidth]{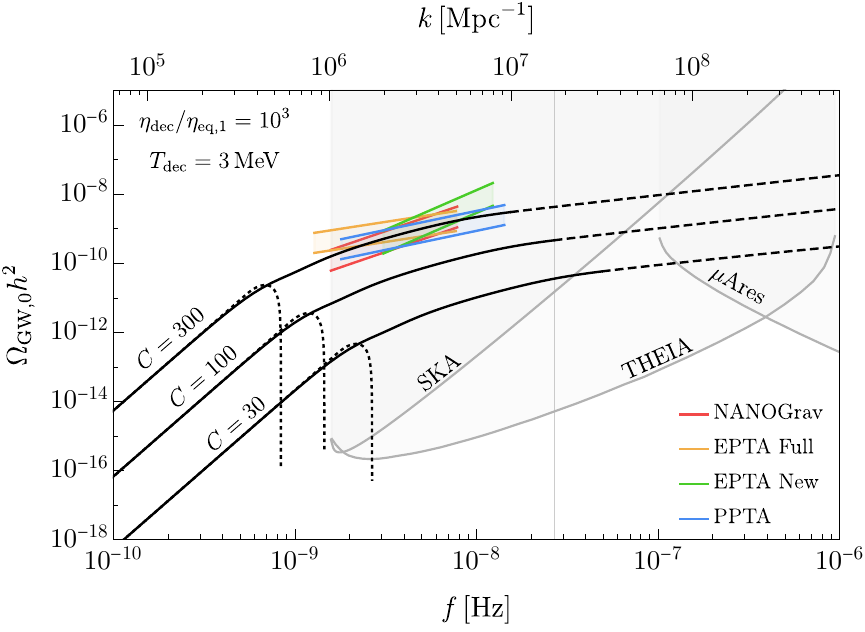}
    \caption{%
        Current density parameter of the gravitational waves for $T_\mathrm{dec} = 3$\,MeV and $C = 30, 100$, and $300$ from bottom to top.
        The dashed and dotted lines are used in the same way as in Fig.~\ref{fig: GW shape}.
        We fix $\eta_\mathrm{dec}/\eta_\mathrm{eq,1} = 1000$.
        The colored bands are the 95\% regions for the best-fit tilt indicated by the recent PTA experiments, NANOGrav~\cite{NANOGrav:2023gor}, EPTA~\cite{Antoniadis:2023ott} with InPTA data, and PPTA~\cite{Reardon:2023gzh}.
        The frequency ranges are determined as the region where the signals seem to be inconsistent with the null hypothesis.
        The gray-shaded regions are the future sensitivities of SKA~\cite{Janssen:2014dka,Weltman:2018zrl}, THEIA~\cite{Garcia-Bellido:2021zgu}, and $\mu$Ares~\cite{Sesana:2019vho}.
        The sensitivity of SKA is taken from Ref.~\cite{Schmitz:2020syl}
        }
    \label{fig: GW vs PTA}
\end{figure}

Finally, we translate the gravitational wave energy density parameter during the RD era to that at the present time by~\cite{Ando:2018qdb}
\begin{align}
    \Omega_\mathrm{GW,0}(k) h^2
    =
    0.83 \left( \frac{g_{*, c}}{10.75} \right)^{-1/3}
    \Omega_{\mathrm{r}, 0} h^2 
    \Omega_{\mathrm{GW,RD}}^{(\mathrm{res})} (\eta_c, k)
    \ ,
\end{align}
where $g_{*, c}$ is the effective degrees of freedom at $\eta_c$, and $\Omega_{\mathrm{r},0}$ is the current density parameter of radiation.
We use $g_{*,c} = 10.75$ and $\Omega_{\mathrm{r},0} h^2 = 4.2 \times 10^{-5}$.
We show the current density parameter of the gravitational waves for $T_\mathrm{dec} = 3$~MeV, $\eta_\mathrm{dec}/\eta_\mathrm{eq,1} = 1000$, and $C = 30, 10, 300$ in Fig.~\ref{fig: GW vs PTA}.
Again, the dotted lines represent the gravitational waves sourced by the scalar perturbations in the linear regime.
The recent PTA data favor the signal for $T_\mathrm{dec} = 3$~MeV, $\eta_\mathrm{dec}/\eta_\mathrm{eq,1} = 1000$, and $C = 300$.
We show the frequency dependence of each factor of $S_\mathrm{low,NL}$ in Appendix~\ref{Appendix}.
Note that these values are not the only parameter set to match the PTA data.
For instance, larger $T_\mathrm{dec}$ and $C$ will also be favored by the data.
In any case, we can see that the non-linear evaluation is crucial to compare the gravitational wave spectrum with the PTA data.

\section{Summary and Discussion}
\label{sec: summary and discussion}

In this paper, we have extended the analysis for the Q-ball scenario in Ref.~\cite{Kasuya:2022cko} and discussed the second-order gravitational waves sourced by the scalar perturbations including the non-linear regime.
By using the fitting formula for the non-linear density perturbations, we could evaluate the gravitational wave spectrum for larger frequencies than previous work.
The resultant gravitational wave spectrum depends on the properties of the Q-ball decay and primordial curvature perturbations.
For a certain parameter set, which can be realized in the Q-ball scenario, the GW spectrum becomes consistent with the signal recently reported by the PTA experiments. 
Moreover, the Q-ball scenario is based on the MSSM framework, and gravitinos are produced after the reheating.
After the dilution by the Q-ball decay, gravitinos can account for dark matter.
In this sense, this scenario can simultaneously explain the primordial helium abundance, gravitational waves suggested by the PTA experiments, and dark matter.

In our scenario, density perturbations grow and become non-linear during the eMD era.
Consequently, microhalos are formed during the eMD era and dissipate at the end of the eMD era.
These phenomena affect the formation and properties of halos formed after the standard matter-radiation equality~\cite{Blanco:2019eij,StenDelos:2019xdk,Barenboim:2021swl,Ganjoo:2023fgg}.
From this point of view, our scenario can also be probed through the small-scale structures.

Before closing the paper, we make some comments on the limitations of our analysis.
Our result is the first step to evaluate the gravitational waves sourced by the scalar perturbations in the non-linear regime, and some improvements can be considered.
First, our analysis adopts the estimate of $S_\mathrm{dec}$ obtained in the linear regime.
In this sense, we expect that a more robust spectrum for a wider frequency range can be obtained with a deeper understanding of the non-linear dynamics through numerical investigation such as $N$-body simulations.

Second, the scalar perturbations in the non-linear regime become non-Gaussian through the growth.
Since our evaluation of the sourced gravitational waves assumes the Gaussian properties of the gravitational potential, its non-Gaussianity will result in some correction in the gravitational wave spectrum.

Third, in this paper, we have assumed that all the Q-balls complete their decay at the same time.
However, the decay process of the Q-balls can deviate in time due to fluctuations in the initial mass or formation time in a realistic situation.
In this case, the universe experiences a more gradual transition from the eMD era to the RD era, which will suppress the gravitational waves as discussed in Ref.~\cite{Inomata:2020lmk}.

Finally, we have focused on the resonant gravitational waves sourced by the primordial curvature perturbations.
The second-order gravitational waves also have non-resonant contributions.
Although the non-resonant contribution is smaller than the resonant contribution around the non-linear scale, $k_\mathrm{cut}$~\cite{Inomata:2019ivs,Inomata:2020lmk}, it will become non-negligible in lower frequency regions.
In addition, we can also consider the density perturbation of Q-balls in analogy to the analysis in the PBH evaporation scenario~\cite{Domenech:2020ssp,Domenech:2021wkk,Bhaumik:2022pil,Papanikolaou:2022chm,Bhaumik:2022zdd,Domenech:2023mqk,Basilakos:2023xof}.
In the PBH scenario, such a contribution produces gravitational waves in a higher frequency region~\cite{Bhaumik:2022pil}, which will also be true in our scenario.
It will be interesting to study the detectability of such gravitational waves in interferometer experiments and the effects of non-linearity on this contribution, but it is out of the scope of this work.

\begin{acknowledgments}
    We would like to thank Keisuke Inomata and Naoki Yoshida for their helpful comments.
    This work was supported by JSPS KAKENHI Grant Nos. 20H05851(M.K.), 21K03567(M.K.), and 23KJ0088(K.M.), and World Premier International Research Center Initiative (WPI Initiative), MEXT, Japan (M.K.).
\end{acknowledgments}

\appendix
\section{Frequency dependence of \texorpdfstring{$S_\mathrm{low,NL}$}{}}
\label{Appendix}

\begin{figure}[t]
    \centering
    \includegraphics[width=.8\textwidth]{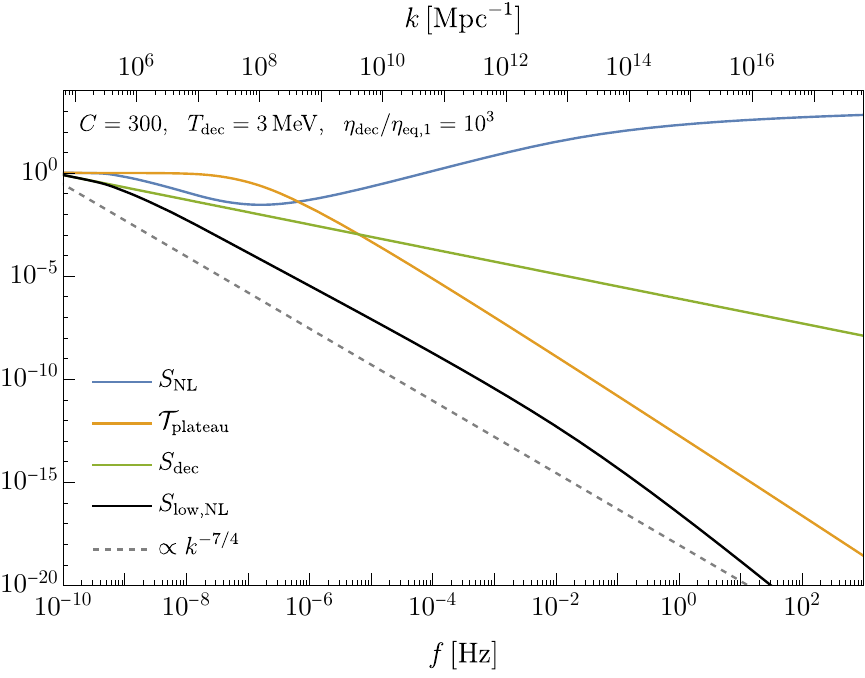}
    \caption{%
        Frequency dependence of $S_\mathrm{low,NL}$ and its ingredients.
        The gray dashed line is proportional to $k^{-7/4}$, which is the dependence of $S_\mathrm{low,NL}$ realizing a flat gravitational wave spectrum (see Eq.~\eqref{eq: 2nd GW formula Mod}).
        }
    \label{fig: SlowNL}
\end{figure}

To understand the effect of $S_\mathrm{low,NL}$ on the shape of the gravitational wave spectrum, we show $S_\mathrm{low,NL}$ and its ingredients, $S_\mathrm{dec}$, $S_\mathrm{NL}$, and $\mathcal{T}_\mathrm{plateau}$, in Fig.~\ref{fig: SlowNL}.
The black solid line shows $S_\mathrm{low,NL}$ while the gray solid line is proportional to $k^{-7/4}$, which corresponds to the dependence of $S_\mathrm{low,NL}$ to realize a flat gravitational wave spectrum.
We can see that $\Omega_{\mathrm{GW},0} h^2$ increases in the frequency range shown in Fig.~\ref{fig: GW vs PTA} even with decreasing $S_\mathrm{low,NL}$.
We can also see that $\Omega_{\mathrm{GW},0} h^2$ decreases for $f \gtrsim \mathcal{O}(10^{-2})$\,Hz although our estimate of $S_\mathrm{NL}$ can be inaccurate for such high frequencies.

\small
\bibliographystyle{JHEP}
\bibliography{Ref}

\end{document}